\documentclass[twocolumn,floatfix,preprintnumbers,citeautoscript,newabstract,aps,superscriptaddress]{revtex4} 
\usepackage{amsmath,amssymb,graphics}
\usepackage[DIV12]{typearea}
\usepackage{amsmath,amsfonts,amssymb}
\usepackage{graphicx}
\usepackage{units}
\usepackage{mathrsfs}
\usepackage{bbm}
\usepackage{pifont}
\usepackage{braket}
\usepackage{units}
\usepackage[caption=false]{subfig} 
\usepackage[dvipsnames]{xcolor}
\usepackage{mathtools}
\usepackage{tikz}
\usepackage[bookmarks=false]{hyperref}
\usepackage{enumerate}
\usepackage[normalem]{ulem}
\usepackage{cancel}
\usepackage{pdfcomment}
\usepackage{xspace}

\newcommand{\Eqref}[1]{Eq.~\eqref{#1}}
\newcommand{\Figref}[1]{FIG.~\ref{#1}}
\newcommand{\Tabref}[1]{Table~\ref{#1}}
\newcommand{\Secref}[1]{Section~\ref{#1}}
\newcommand{\Appref}[1]{Appendix~\ref{#1}}

\newcommand{\eVdist}{\kern-0.06em}

\newcommand*{\rep}[2][]{\ensuremath{{\boldsymbol{#2}#1}}}

\newcommand{\CenterObject}[1]{\ensuremath{\vcenter{\hbox{#1}}}}

\newcommand{\I}{\mathrm{i}}

\newcommand{\Z}[1]{\ensuremath{\mathbbm{Z}_{#1}}} 
\newcommand{\Id}{\mathbbm{1}}

\newcommand{\Dff}{\ensuremath{\Delta(54)}\xspace}

\newcommand{\CP}{\ensuremath{\mathcal{CP}}\xspace}

\definecolor{darkgreen}{rgb}{0.0, 0.6, 0.2}

\allowdisplaybreaks[1]

\title{\textbf{\boldmath \CP Violation from String Theory\unboldmath}}

\begin{document}

\preprint{UCI-2018-07; TUM-HEP 1156/18\\}

\title{\textbf{\boldmath \CP Violation from String Theory\unboldmath}}

\author{Hans Peter Nilles}
\email[]{nilles@th.physik.uni-bonn.de}
\affiliation{Bethe Center for Theoretical Physics und Physikalisches
Institut der Universit\"at Bonn, Nussallee 12, 53115 Bonn, Germany}
\author{Michael Ratz}
\email[]{mratz@uci.edu}
\affiliation{Department of Physics and Astronomy, 
University of California, Irvine, California 92697-4575, USA}
\author{Andreas Trautner}
\email[]{atrautner@uni-bonn.de}
\affiliation{Bethe Center for Theoretical Physics und Physikalisches 
Institut der Universit\"at Bonn, Nussallee 12, 53115 Bonn, Germany}
\author{Patrick K.S.\ Vaudrevange}
\email[]{patrick.vaudrevange@tum.de}
\affiliation{Physik Department T75, 
Technische Universit\"at M\"unchen, James-Franck-Stra\ss e~1, 
85748 Garching, Germany}
%


\begin{abstract}
We identify a natural way to embed \CP symmetry and its violation
in string theory. The \CP symmetry of the low energy effective theory
is broken by the presence of heavy string modes. \CP violation is
the result of an interplay of \CP and flavor symmetry. \CP violating
decays of the heavy modes could originate a cosmological
matter-antimatter asymmetry.
\end{abstract}

\maketitle

\section{Introduction}
Aspects of \CP symmetry and its violation play a crucial role in several physics
phenomena. This includes the question of \CP symmetry in strong interactions
(the so-called strong \CP problem), the violation of \CP in the Yukawa sector of
the standard model (SM) (with at least 3 families of quarks and leptons) and the
desire for a source of \CP violation (CPV) in the process of a dynamical creation of
the cosmological matter-antimatter asymmetry. We are thus confronted with the
following questions:
What is the origin of \CP symmetry and its violation? Is there a relation to the
flavor symmetries in the SM of particle physics? Is there a
``theory of \CP'' in the ultraviolet completion of the SM that
explains both the origin of \CP symmetry and its breakdown?

In the present letter we try to address these questions about \CP and flavor
symmetries in the framework of string theory. Our approach to a ``theory of \CP''
is based on orbifold compactifications of heterotic string theory (the
so-called MiniLandscape \cite{Lebedev:2006kn,Lebedev:2007hv,Lebedev:2008un,Nilles:2014owa}) but 
should be valid qualitatively for a wide range of
string theory constructions. From our exploration of these models the following
general picture emerges:

\begin{itemize}
\item we find \CP candidates strongly connected to flavor symmetries,
specifically \CP as an outer automorphism of the flavor group;
\item the light (``massless'') string spectrum results in a low-energy
effective field theory with a well-defined \CP transformation, 
which can be conserved only in the absence of couplings to the heavy modes;
\item the presence of heavy modes (here the winding modes of string theory)
initiates a breakdown of \CP (similar to the picture of ``explicit geometrical 
\CP violation'');
\item \CP violating decays of the heavy (winding) modes could induce the
cosmological matter-antimatter asymmetry. Other possible CPV effects
can be induced through couplings of light fields to the heavy modes.
\end{itemize}

This provides us with a picture where the source of \CP breakdown is 
already included within the construction of the symmetry itself. 
It also shows that the breakdown of \CP requires a certain amount of 
complexity of the theory (reminiscent of the need of three families 
in the CKM case).

The origin of \CP violation in the context of string theory and extra 
dimensions has been discussed in many regards, see \cite{Ibrahim:2007fb} 
for a review and references therein. Our approach is new in the following 
sense: While it has been known that extra dimensions provide an origin of 
discrete (flavor) symmetries \cite{Kobayashi:2006wq,Nilles:2012cy,Beye:2014nxa},
a more recent insight, based on the original idea of ``explicit geometrical \CP 
violation'' \cite{Chen:2009gf}, is that a large class of discrete groups is  
generally incompatible with \CP \cite{Chen:2014tpa}. 
This comes about because these groups do not allow for 
complex conjugation outer automorphisms which, however, correspond to physical 
\CP transformations in the most general sense \cite{Holthausen:2012dk, Trautner:2016ezn}.
In these cases, \CP is explicitly violated by phases which are discrete 
and calculable because they originate from the complex Clebsch-Gordan 
coefficients of the respective flavor group.
The main progress in this letter is to demonstrate that such a situation 
arises naturally in string theory.

As a specific example we consider a $\Z3$ orbifold with flavor group \Dff that appears naturally in 
the MiniLandscape constructions \cite{Kobayashi:2006wq}. In this case, \CP should be a
subgroup of $\mathrm{S}_4$, the group of outer automorphisms of \Dff; thus
flavor group and \CP are intimately related. The irreducible representations of
\Dff include singlets, doublets, triplets and anti-triplets.
The massless spectrum of the theory, however, contains only singlets and
triplets (as well as anti-triplets) of \Dff. This allows for a
\CP symmetric low-energy effective field theory of the massless states. The
presence of the heavy winding modes that transform as doublets of \Dff
leads to an obstruction for the definition of \CP symmetry 
thereby realizing the mechanism of ``explicit geometrical \CP violation''. 
All \CP violating effects originate
through couplings of the light states to at least three non-trivial doublets.
\CP violating decays of the heavy doublets
are a generic property of the scheme. Combined with baryon- and/or
lepton-number violation this could lead to a cosmological baryon- and/or lepton
asymmetry.

\section{$\boldsymbol{\Dff}$ Flavor Symmetry from String Theory and the Light Spectrum}
\label{sec:DiscreteSymmetries}

In order to understand the origin of \Dff from strings it is 
sufficient to concentrate on the compactification of two extra dimensions 
on a $\mathbbm{T}^2/\Z{3}$ orbifold. For a full string model this 
$\mathbbm{T}^2/\Z{3}$ can easily be extended to a six-dimensional orbifold, 
e.g.\ $\mathbbm{T}^6/\Z{3}\times\Z{3}$.

\begin{figure}[t]
 \includegraphics[width=1\linewidth]{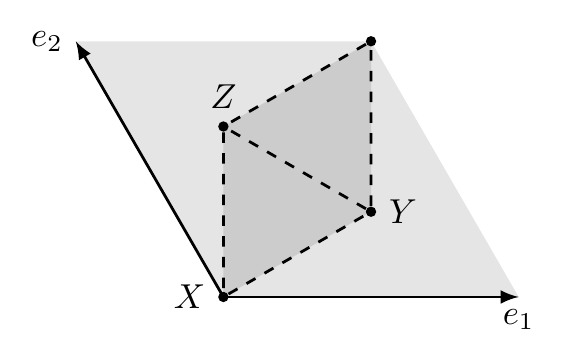}
 \caption{\label{fig:Orbifold} $\mathbbm{T}^2/\Z3$ orbifold with fixed 
 points $X$, $Y$, $Z$. The fundamental domain is shaded in dark gray.}
\end{figure}

Geometrically, a $\mathbbm{T}^2/\Z{3}$ orbifold can be defined in two
steps: (i) one defines a torus $\mathbbm{T}^2$ by specifying a 
lattice $\Lambda = \{n_1 e_1 + n_2 e_2 \,|\, n_i \in \Z{}\}$, spanned
by the vectors $e_1$ and $e_2$. We choose $|e_1| = |e_2|$ and the 
angle between $e_1$ and $e_2$ is set to $120^\circ$. (ii) one 
identifies points on $\mathbbm{T}^2$ that differ by a 
120$^\circ$ rotation generated by $\theta$. The resulting orbifold 
has the shape of a triangular pillow, see \Figref{fig:Orbifold} and~\ref{fig:WindingModes}.

Closed strings on the $\mathbbm{T}^2/\Z{3}$ orbifold come in
three classes: (i) trivially closed strings, which are closed even in 
uncompactified space, (ii) winding strings with winding numbers $n_1$ 
and $n_2$ in the torus directions $e_1$ and $e_2$, 
respectively, and (iii) twisted strings, which are closed only up to a 
$\theta^k$ rotation for $k=1,2$. For $k=1$ or $k=2$ they belong to the 
so-called first or second twisted sector, respectively. On the other 
hand, trivially closed strings and winding strings belong to the 
so-called untwisted sector and live in the bulk of the orbifold. 
In contrast, twisted strings are localized at the three corners 
(fixed points) of the $\mathbbm{T}^2/\Z{3}$ orbifold. For $k=1,2$, 
they are created by twisted vertex operators which we label as
\begin{equation}\label{eqn:twistedstrings}
\chi^{(\rep[_1]{3})} \sim \left(\!\!\begin{array}{c}
X \\ Y \\ Z 
\end{array}\!\!\right),  \quad\text{and}\quad
\psi^{(\rep[_1]{\bar3})} \sim \left(\!\!\begin{array}{c}
\bar{X} \\ \bar{Y} \\ \bar{Z} 
\end{array}\!\!\right)\;,
\end{equation}
respectively (compared to Ref.~\cite{Lauer:1989ax} we set 
$X = \sigma_0^+$, $Y = \sigma_1^+$, $Z = \sigma_2^+$ and 
$\bar{X} = \sigma_0^-$, $\bar{Y} = \sigma_1^-$, 
$\bar{Z} = \sigma_2^-$).

Interactions of strings on orbifolds are restricted by selection 
rules. In the case of $\mathbbm{T}^2/\Z{3}$ the point group (PG) 
and space group (SG) selection rules result in a 
$\Z{3}^\text{PG}\times\Z{3}^\text{SG}$ symmetry \cite{Hamidi:1986vh}.
Massless untwisted strings transform trivially, while twisted 
strings transform as
\begin{subequations}\label{eq:PGSG}
\begin{eqnarray}
\chi^{(\rep[_1]{3})} & \xmapsto{~\Z{3}^\text{PG}~} & \mathrm{diag}(\omega,\omega,\omega)\; \chi^{(\rep[_1]{3})}\;, \\
\chi^{(\rep[_1]{3})} & \xmapsto{~\Z{3}^\text{SG}~} & \mathrm{diag}(1,\omega,\omega^2)\; \chi^{(\rep[_1]{3})}\;, \\
\psi^{(\rep[_1]{\bar3})} & \xmapsto{~\Z{3}^\text{PG}~} & \mathrm{diag}(\omega^2,\omega^2,\omega^2)\; \psi^{(\rep[_1]{\bar3})}\;, \\
\psi^{(\rep[_1]{\bar3})} & \xmapsto{~\Z{3}^\text{SG}~} & \mathrm{diag}(1,\omega^2,\omega)\;\; \psi^{(\rep[_1]{\bar3})}\;,
\end{eqnarray}
\end{subequations}
where $\omega:=\mathrm{e}^{2\pi\I/3}$. 
In the absence of non-trivial backgrounds on $\mathbbm{T}^2/\Z{3}$
there is in addition an $S_3$ symmetry corresponding to all 
permutations of the three twisted strings. Combining this symmetry 
with the PG and SG symmetries, one obtains a \Dff flavor symmetry,
see \Appref{app:delta54}. Massless untwisted strings transform as
trivial singlets $\rep[_0]{1}$, while the twisted strings 
$\chi^{(\rep[_1]{3})}$ and $\psi^{(\rep{\bar{3}}_1)}$ transform 
as $\rep[_1]{3}$ and $\rep[_1]{\bar3}$ of \Dff, 
respectively \cite{Kobayashi:2006wq, Beye:2014nxa}.

\section{$\boldsymbol{\Dff}$ and Explicit Geometrical $\boldsymbol{\CP}$ Violation}
\label{sec:GeometricCPV}

Let us discuss some details of \Dff and how this group can lead to
the phenomenon of explicit geometrical \CP violation.
The non-trivial irreps of \Dff are the real \rep[_1]{1}, 
a quadruplet of real doublets \rep[_{k=1,2,3,4}]{2} as well as the 
faithful complex triplets \rep[_1]{3}, \rep[_2]{3} and their 
respective complex conjugates \rep[_1]{\bar3} and \rep[_2]{\bar3}. 
Tensor products relevant to this work are
\begin{subequations}
\begin{align}\label{eq:3timesbar3}
 \rep[_i]{3}\otimes\rep[_i]{\bar3}\,&=\rep[_0]{1}\oplus\rep[_1]{2}
 \oplus\rep[_2]{2}\oplus\rep[_3]{2}\oplus\rep[_4]{2}\;,&\\
 \rep[_k]{2}\otimes\rep[_k]{2}&=\rep[_0]{1}\oplus\rep[_1]{1}
 \oplus\rep[_k]{2}\;.&\label{eq:2times2}
\end{align}
\end{subequations}
The outer automorphism group (Out) of \Dff is
\begin{equation}
 \mathrm{Out}\left[\Dff\right]\cong\mathrm{S}_4\;,
\end{equation}
the permutation group of four elements. 
On the four doublets, $\mathrm{S}_4$ acts as all possible 
permutations. On the triplets, odd permutations 
in $\mathrm{S}_4$ act as complex conjugation, while even 
permutations in $\mathrm{S}_4$ map the triplets to themselves 
\cite{Trautner:2016ezn}. 
In addition, all these transformations are typically endowed 
with matrices that act on the representations internally.

A physical \CP transformation maps all fields of a theory
in some irreps \rep{r} to their respective complex conjugate 
fields, which transform in $\rep{r}^*$ \cite{Trautner:2016ezn}.
Therefore, a physical \CP transformation should be an outer 
automorphism of \Dff which maps all occurring irreps to their 
respective complex conjugates \cite{Holthausen:2012dk, Chen:2014tpa}.
However, depending on the specific group, such outer automorphisms 
do not need to exist, and groups which do not have them
are called ``type I'' \cite{Chen:2014tpa}.

It turns out that \Dff is a group of type I. This becomes manifest 
by the fact that in the presence of triplets 
it is \textit{only} possible to find a physical \CP 
transformation as subset of $\mathrm{Out}\left[\Dff\right]$, 
\textit{if} a given theory contains fields in \textit{no more than two}
distinct doublet representations. By contrast, \textit{if} a given 
theory with triplets contains \textit{more than two} doublets, physical \CP is 
violated by complex Clebsch-Gordan coefficients of \Dff and this is
called explicit geometrical \CP violation.

Coming back to our model, we recall that the light (from a string 
perspective massless) spectrum consists only of \Dff triplets 
and singlets. Consequently, there exist outer automorphisms of $\Dff$ 
which correspond to physical \CP transformations for the light 
spectrum. One may thus be led to the conclusion that \CP can be 
conserved in this model even though $\Dff$ is a group of type I.
However, this conclusion is premature because it disregards 
heavy string modes. We will take them into account in the next section.

\section{$\boldsymbol{\Dff}$ Doublets from Heavy Winding States}
\label{sec:WindingModes}

\begin{figure*}[ht!]
\subfloat[]{\label{fig:WindingModes1}
\centering\includegraphics[width=0.245\linewidth]{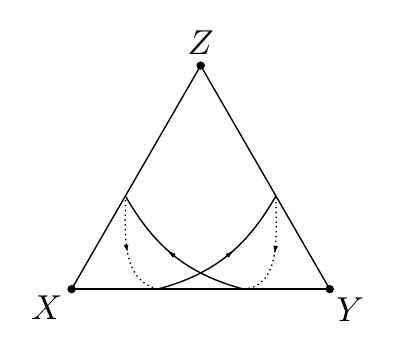}}
\subfloat[]{\label{fig:WindingModes2}
\centering\includegraphics[width=0.245\linewidth]{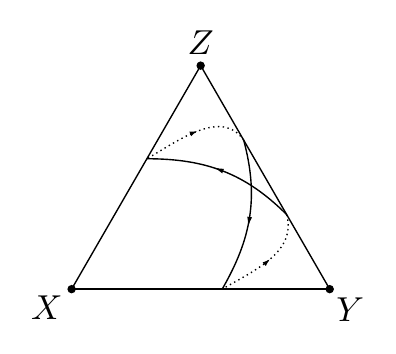}}
\subfloat[]{\label{fig:WindingModes3}
\centering\includegraphics[width=0.245\linewidth]{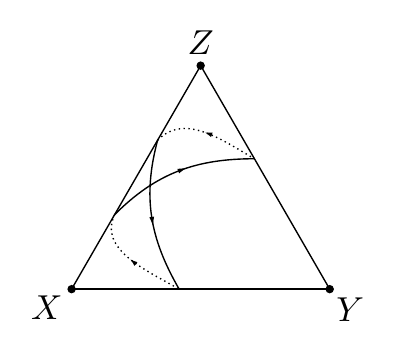}}
\subfloat[]{\label{fig:WindingModes4}
\centering\includegraphics[width=0.245\linewidth]{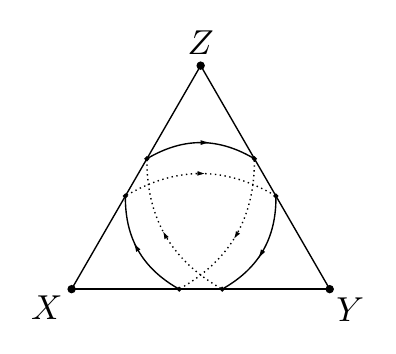}}
\caption{\label{fig:WindingModes}Illustration of the geometrical 
winding strings. For example, (a), (b), and (c) depict the winding modes 
from $X\bar{Y}$, $Y\bar{Z}$, and $Z\bar{X}$ which all have winding number $N=2$.}
\end{figure*}

An untwisted string on $\mathbbm{T}^2/\Z{3}$ is generally massless 
only if it does not wind around the torus $\mathbbm{T}^2$ and does 
not carry Kaluza-Klein (KK) momentum. For this reason, winding strings often 
have been ignored in the literature. 
In this respect, one of our main results is that general untwisted strings 
on $\mathbbm{T}^2/\Z{3}$ can transform as doublets of \Dff, 
as we show in the following.

A general untwisted string with vertex operator $V^{p,w}$ is 
characterized by its winding on the lattice, $w \in \Lambda$, 
and momentum on the dual lattice, $p \in \Lambda^*$. It can be 
constructed by joining two strings: one string from the first 
twisted sector $X$, $Y$, $Z$ combines with another string from 
the second twisted sector $\bar{X}$, $\bar{Y}$, $\bar{Z}$. 
Such processes are described by the corresponding operator product 
expansions (OPEs), which are given explicitly for 
$\mathbbm{T}^2/\Z{3}$ in Ref.~\cite{Lauer:1989ax}. 
Solving these OPEs for the untwisted strings we get
\begingroup
\renewcommand*{\arraystretch}{1}
\arraycolsep=0pt
\thinmuskip=1mu
\medmuskip=2mu
\thickmuskip=1mu
\newcommand{\hminus}{\hspace{-6pt}}
\begin{subequations}\label{eqn:InvertedOPEs}
\begin{align}
&\rep[_0]{1}\!:& V^{(00)}&=\frac{1}{3}
\left(X\bar{X} + Y\bar{Y} + Z\bar{Z} \right)\;,& \\ 
&\rep[_1]{2}\!:& \hminus
\left(
\begin{array}{c}
V^{(02)} \\
V^{(01)}
\end{array}\right)&=
\frac{1}{3}\left(
\begin{array}{c}
Y\bar{Z} + Z\bar{X} + X\bar{Y}\\
Z\bar{Y} + X\bar{Z} + Y\bar{X}
\end{array}\right),& \raisetag{20pt} \\ 
&\rep[_2]{2}\!:& \hminus
\left(
\begin{array}{c}
V^{(10)} \\
V^{(20)} 
\end{array}\right)&=
\frac{1}{3}\left(\begin{array}{c}
X\bar{X} + \omega^{\phantom{2}}\,   Y\bar{Y} + \omega^2\, Z\bar{Z} \\
X\bar{X} + \omega^2\, Y\bar{Y} + \omega^{\phantom{2}}\,   Z\bar{Z}
\end{array}\right),& \raisetag{20pt} \\
&\rep[_3]{2}\!:&  \hminus
\left(\begin{array}{c}
V^{(12)} \\
V^{(21)}
\end{array}\right)&=
\frac{1}{3}\left(\begin{array}{c}
Y\bar{Z} + \omega^{\phantom{2}}\,   Z\bar{X} + \omega^2\, X\bar{Y}\\
Z\bar{Y} + \omega^2\, X\bar{Z} + \omega^{\phantom{2}}\,   Y\bar{X}
\end{array}\right),& \raisetag{20pt} \\
&\rep[_4]{2}\!:&  \hminus
\left(\begin{array}{c}
V^{(11)} \\
V^{(22)}
\end{array}\right)&=
\frac{1}{3}\left(\begin{array}{c}
Z\bar{Y} + \omega^{\phantom{2}}\,   X\bar{Z} + \omega^2\, Y\bar{X}\\
Y\bar{Z} + \omega^2\, Z\bar{X} + \omega^{\phantom{2}}\,   X\bar{Y}
\end{array}\right).& \raisetag{20pt}
\end{align}
\end{subequations}
\endgroup
Here we have introduced classes of untwisted strings%
\begin{equation}
\!\!V^{(MN)}\!:=\!\sum_{\substack{\;p \in \Lambda_M^* \\ w\in \Lambda_N}} 
\!\!\! C_{p,w} V^{p,w} ~\;\text{for}\;~ M,N = 0,1,2\;,
\end{equation}%
where $\Lambda_N$ is a sublattice of $\Lambda$ with winding number
$N := n_1 + n_2$ mod $3$ and $\Lambda_M^*$ the sublattice of 
$\Lambda^*$ with KK number $M :=-m_1 + m_2$ mod $3$ (Note the 
difference to~\cite{Lauer:1989ax} due to a basis change of the 
torus lattice $\Lambda$). Furthermore, the coefficients $C_{p,w}$ are
defined in Ref.~\cite{Lauer:1989ax} and they tend to zero, 
$C_{p,w} \rightarrow 0$, for several limits: for higher windings 
$|w| \rightarrow \infty$, for higher momenta $|p|\rightarrow \infty$ or, 
keeping $p$ and $w$ fixed, for larger torus radii $|e_1| = |e_2|\rightarrow \infty$.

Comparing Eqs.~\eqref{eqn:InvertedOPEs} to the \Dff tensor product 
$\rep[_1]{3}\otimes\rep[_1]{\bar3}$, we identify the multiplets of 
winding strings as $\rep[_0]{1}$, $\rep[_1]{2}$, $\rep[_2]{2}$, 
$\rep[_3]{2}$ and $\rep[_4]{2}$, respectively, see \Eqref{eq:3timesbar3} 
and \Eqref{eq:Delta54CGs}. The \Dff doublets $\rep[_k]{2}$ of winding 
strings are generally massive (where the mass terms are \Dff invariant, see 
\Eqref{eq:2times2}) and their masses are in general different.

There is a simple geometric intuition for these findings, revealing a 
remarkable difference between the winding modes in irreps
$\rep[_{1,3,4}]{2}$, as compared to the modes in $\rep[_2]{2}$ and 
$\rep[_0]{1}$. The classes of untwisted strings $V^{(M N)}$
form \Dff covariant combinations of certain ``geometric'' winding 
modes, e.g.\ $X\bar{Y}$. For $\rep[_{1,3,4}]{2}$ these modes wind 
once around one fixed point and in a different orientation around 
another, see \Figref{fig:WindingModes1}-\ref{fig:WindingModes3}.
The two different components of the doublets wind in opposing 
directions. By contrast, the doublet $\rep[_2]{2}$ is formed by a 
geometrical winding mode that has net zero winding number around all
fixed points, see \Figref{fig:WindingModes4}.

\section{$\boldsymbol{\CP}$ Violation from Heavy Winding Modes}
\label{sec:OriginOfCPV}

The effective operators in the holomorphic superpotential are given 
by $\Delta(54)$ invariants. For example, the simplest direct 
couplings between representations $\rep{3}_1$ and $\rep{\bar{3}}_1$, 
from the first and second twisted sector, respectively, 
to heavy winding modes 
in representations $\rep{2}_k$ are given by contractions of the form
\begin{equation}\label{eq:coupling}
 \mathscr{W}\supset\sum_k \left(c_k\right)^{mab} 
 \phi^{(\rep{2}_k)}_m \chi^{(\rep{3}_1)}_a 
 \psi^{(\rep{\bar{3}}_1)}_b \;.
\end{equation}
Here we have introduced exemplary fields 
$\phi^{(\rep{2}_k)}\equiv\phi_k$, $\chi^{(\rep{3}_1)}$, and 
$\psi^{(\rep{\bar{3}}_1)}$ in the according representations. 
The sum over the internal $\Delta(54)$ doublet and triplet components,
denoted by $a,b=1,2,3$ and $m=1,2$, is implicit. 
The coupling tensors to different winding modes 
$\left(c_k\right)^{mab}$ ($k=1,2,3,4$) are fixed by the requirement 
of $\Delta(54)$ invariance up to a global normalization $|c_k|$, 
corresponding to the overall coupling strength which is determined by 
the $T$ modulus (e.g.\ by the size of the $\mathbbm{T}^2/\Z{3}$ orbifold). 
The explicit form of the coupling tensors $c_k$ is obtained 
from the Clebsch-Gordan coefficients (cf.\ \Eqref{eq:Delta54CGs} in 
\Appref{app:delta54}) in a straightforward way.
The presence of these couplings, i.e.\ $c_k\neq0$ $\forall k$, 
is inconsistent with $\textit{any}$ physical \CP transformation
\begin{equation}\label{eq:CPtrafo}
 \phi_k\mapsto U_k\phi_k^*\;,~~ \chi\mapsto U_\chi\chi^*\;,~~ 
 \text{and}~~ \psi\mapsto U_\psi\psi^*\;,
\end{equation}
where we have allowed for the most general form of this 
transformation with arbitrary unitary matrices $U_{k,\chi,\psi}$ 
while we have suppressed the transformation of the space-time 
argument. Therefore, \CP is explicitly violated by the couplings 
\Eqref{eq:coupling}.

As argued in \Secref{sec:GeometricCPV}, this can readily be understood 
directly from group theory. One may also check here that there is no basis 
in which all coupling tensors $c_{k=1,2,3,4}$ are simultaneously real 
(which, however, is not sufficient to claim CPV, as complex couplings
can co-exist with \CP conservation if there are conserved
higher order \CP transformations \cite{Chen:2014tpa}\footnote{%
Other examples of higher order CP transformations with irreducible 
complex coupling coefficients exist in the context of multi-Higgs 
doublet models \cite{Ivanov:2015mwl} and in the context of extra 
dimensions \cite{Chang:2001yn,Chang:2001uk} 
(we have to stress, however, that our notion of
geometrical CPV is substantially different from the same-named phenomenon 
in the latter references).}).

\begin{figure}[t!]
 \CenterObject{
 \includegraphics[width=0.8\linewidth]{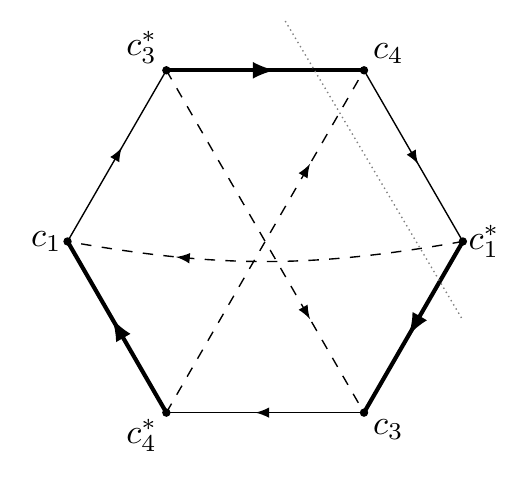}}
 \caption{\label{fig:Invariants}%
 The lowest-order complex (\CP-odd) basis invariant that can be 
 formed out of the \Dff invariant coupling tensors $c_{k=1,3,4}$ of 
 \Eqref{eq:coupling}. Dashed lines correspond to $\Delta(54)$ 
 doublets $\phi$, thin solid lines correspond to the $\Delta(54)$ 
 triplets $\chi$ and thick solid lines correspond to the 
 $\Delta(54)$ anti-triplets $\psi$. 
 Arrows here denote the $\Delta(54)$ charge flow.
 The dotted gray line denotes a cut that gives rise
 to the diagrams in \Figref{fig:decay}.}
\end{figure}

Finally, it is always possible to state physical \CP violation in a 
basis independent way. \CP-odd basis invariants can readily be 
constructed by contracting coupling tensors in such a way that 
(unitary) basis transformations cancel amongst the various index 
contractions. To produce a complex-valued (thus, \CP-odd) basis 
invariant it follows from our previous discussion that \textit{at 
least three different types} of doublets (e.g.\ take $\rep{2}_{1,3,4}$) 
must be involved in such a contraction. This is confirmed by a scan 
over all possible basis invariant contractions of coupling tensors.
The lowest order \CP-odd invariants arise at the four-loop level, 
with an example being displayed in \Figref{fig:Invariants}.
Explicitly, this basis invariant is given by
\begin{eqnarray}
 \mathcal{I}_1 \!\!& = &\!\! \left(c_1\right)^{mab} \!\left(c^*_4\right)^{ncb}\! \left(c_3\right)^{pcd}\! \left(c^*_1\right)^{med}\! \left(c_4\right)^{nef} \!\left(c_3^*\right)^{paf} \nonumber\\
  & = & \frac{1+3\,\mathrm{e}^{4\pi\I/3}}{36} |c_1|^2\,|c_3|^2\,|c_4|^2\;.
\end{eqnarray}
Here the summation over repeated indices is understood and we have 
used $|c_k|^2$ to denote the moduli of the coupling tensors. 
The fixed complex phase of the invariant 
is a group theoretically predicted parameter-independent \CP 
violating (weak) phase. Another \CP-odd invariant of the same order 
can be obtained from $\mathcal{I}_1$ by hermitean conjugation 
[or changing the order of indices of the coupling tensors according 
to $(143)\rightarrow(134)$]. Analogous invariants exist for 
all other sets of three distinct doublets. 
Further \CP-odd invariants exist for couplings of the type 
$\rep[_k]2\otimes\rep3\otimes\rep3\otimes\rep3$, where again, 
at least three different doublets have to be involved in a given 
invariant in order to generate \CP-odd contributions.

\begin{figure}[t!]
\subfloat[]{\label{fig:decay1}
\centering\includegraphics[width=0.6\linewidth]{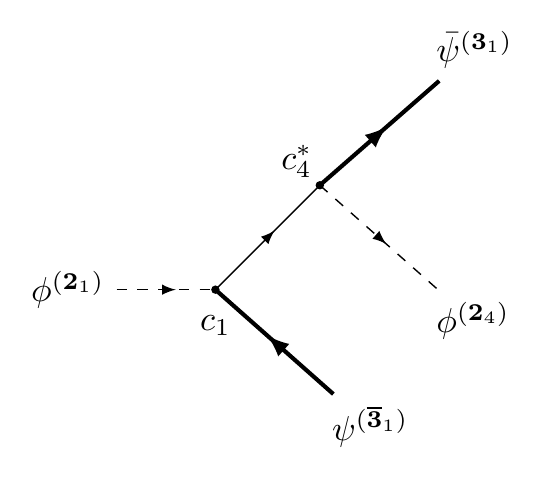}}\\
\subfloat[]{\label{fig:decay2}
\centering\includegraphics[width=0.9\linewidth]{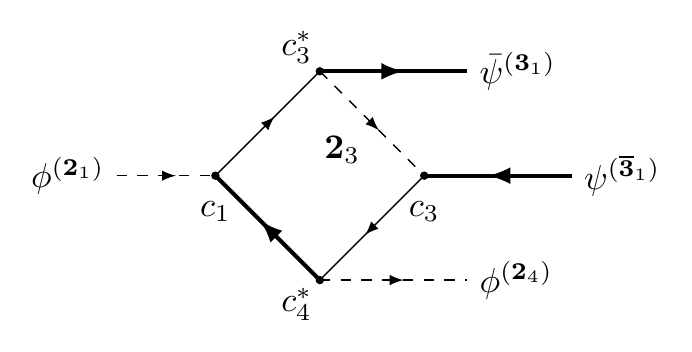}}
\caption{\label{fig:decay}Example tree-level and one-loop diagrams whose interference can give rise to 
\CP violation in a decay $\rep[_1]{2}\to\rep[_1]{\bar{3}}\rep[_1]{3}\rep[_4]{2}$.}
\end{figure}

Diagrams corresponding to \CP violating physical processes such as 
oscillations and/or decays can be obtained from invariants such as the one in
\Figref{fig:Invariants} by cutting edges appropriately. For example, 
the cut indicated by the dotted gray line in \Figref{fig:Invariants}
gives rise to the pair of diagrams in \Figref{fig:decay} 
whose interference generates a \CP asymmetry in
a decay $\rep[_1]{2}\to\rep[_1]{\bar{3}}\rep[_1]{3}\rep[_4]{2}$ \footnote{%
For this decay to be effective it is required that $m_{\rep[_1]{2}}>m_{\rep[_3]{2}},m_{\rep[_4]{2}}$.
Similar decays exist for all possible mass orderings of the doublets.}.

The discussion becomes exceedingly model dependent at this point.
For example, the questions of whether a given decay is (kinematically) allowed, 
whether or not significant \CP asymmetry is generated (individual 
amplitudes can vanish or cancel against one another), or whether 
or not other quantum numbers such as lepton and/or baryon number are violated 
can only be answered in a concrete model.
In existing, semi-realistic string theory models with \Dff flavor symmetry,
SM quarks and leptons as well as flavons transform as \Dff triplets and anti-triplets 
\cite{Carballo-Perez:2016ooy, Ramos-Sanchez:2017lmj}.
For instance, inspecting Tab.\ 1 of \cite{Carballo-Perez:2016ooy} one finds right-handed 
neutrinos in \rep[_1]{\bar{3}} as well as SM neutral flavons in \rep[_1]{3}. 
The \CP and lepton number violating decay of a heavy doublet to final states which include 
these modes, hence, could generate a lepton asymmetry \footnote{%
The decay in \Figref{fig:decay} does not allow to generate a lepton number asymmetry
among the final-state triplets. However, other cuts of the invariant $\mathcal{I}_1$ and of 
the other CP-odd invariants allow for such processes.}.
This illustrates how baryon and/or lepton number asymmetries can be generated by our mechanism.
We do not further detail this discussion for the scope of the present letter since 
we have achieved to establish our main point which is the general existence of group 
theoretical \CP violation in string theory.

\section{Discussion}

We have seen that the mechanism of 
explicit geometrical \CP violation 
has a natural embedding in string theory. 
\CP violation here is for a symmetry reason, enforced by a
discrete (flavor) symmetry of ``type I'' which prohibits any 
physical \CP transformation and dictates discrete \CP violating phases.

In our example, the low-energy effective
theory allows for a \CP transformation, 
but \CP is broken in the presence of heavy
string modes. A generic property of this scheme is the \CP violating
decay of heavy modes that could originate a cosmological
baryon/lepton asymmetry. The discussion of other \CP violating effects
such as $\theta_{\mathrm{QCD}}$ and the CKM phases is strongly model
dependent. As explicit model building is very complicated in
string theory, one would be encouraged to tackle these more
refined questions in bottom-up constructions of the scheme in
detail. This could then help to identify those models that
are of phenomenological interest, small $\theta_{\mathrm{QCD}}$ and realistic
CKM phases. One would hope to reproduce these models in a top-down
construction and understand these specific properties from the
geometry of extra dimensions and the geographical location of
strings in compactified space. The complexity of \CP violation in
the standard model (with 3 families of quarks and leptons) could
then be related to the complexity to the spectrum and couplings
of heavy string states 
(here at least 3 doublets of string winding modes).

From the more theoretical (top-down) point of view there remain
a few question that have to be analyzed in detail. The work of
Ref.\ \cite{Lauer:1989ax,Lauer:1990tm} 
[that lead to the selection rules in \Eqref{eqn:InvertedOPEs}] 
studied the role of T-duality in the framework of compactified string theory.
In a UV-complete picture one now would have to understand
the connection between the flavor group [\Dff in our
example] and T-duality. Especially, whether this does 
lead to a symmetry enhancement of the flavor group and how 
T-duality correlates with the outer automorphisms of $\Dff$ (and thus \CP).

Among other things this would then allow us to decide whether
\CP violation in the UV-complete theory is explicit or spontaneous,
and whether \CP should be interpreted as a discrete gauge symmetry 
\cite{Dine:1992ya,Choi:1992xp}.
Our discussion up to now did not address this question 
as we have only considered the low-energy effective theory 
where the appearing flavor symmetry
is incompatible with \CP, which hence appears to be violated explicitly.
Nevertheless, the fact that \CP violation of this type can originate 
spontaneously, even though in a subtle manner, has been 
previously demonstrated \cite{Ratz:2016scn}.
We would have to consider a UV-complete theory with additional symmetries
(like T-duality) before we could hope for a definite answer here.
A UV-complete theory might as well give further insight
into the phenomenological properties of the scheme, as e.g.\ the
suppression of $\theta_{\mathrm{QCD}}$ or how the remaining flavor symmetry 
is ultimately broken to a realistic pattern.
We hope to report on progress towards
the UV-complete picture in the near future.

As a final remark, we stress that many phenomenologically viable string compactifications 
feature discrete groups of type I, see e.g.\ \cite{Ramos-Sanchez:2017lmj,Ramos-Sanchez:2018xx}.
If all possible irreducible representations of these symmetry groups actually appear in the spectrum 
(fulfilling some kind of stringy ``completeness conjecture'')
all these models have the presented mechanism of \CP violation built in.

\begin{acknowledgments}
P.V.\ is supported by the Deutsche Forschungsgemeinschaft (SFB1258).
The work of H.P.N.\ and A.T.\ has been supported by the German
Science Foundation (DFG) within the SFB-Transregio TR33 ``The Dark Universe''.
The work of M.R.\ is supported by NSF Grant No.\ PHY-1620638 and PHY-1719438. 
M.R. would like to thank the Aspen Center for physics for hospitality and support.
\end{acknowledgments}

\appendix
\section{Conventions for $\boldsymbol{\Dff}$}
\label{app:delta54}

We follow the conventions of Ref.\ \cite{Trautner:2016ezn} where more
details about \Dff can be found 
(cf.\ also \cite{Ishimori:2010au,Escobar:2011mq} but mind the 
different notational conventions). A minimal generating set of 
matrices for the three-dimensional 
representation $\rep{3}_1$ of $\Delta(54)$ is
\begin{equation}\label{eq:Delta54Generators3}
\!\!\!\!A \!=\!\! \begin{pmatrix}
0 & 1 & 0\\
0 & 0 & 1\\
1 & 0 & 0\\
\end{pmatrix}\!\!,
B \!=\!\! \begin{pmatrix}
1 & 0      & 0\\
0 & \omega & 0\\
0 & 0      & \omega^2 \\
\end{pmatrix}\!\!,
C \!=\!\! \begin{pmatrix}
1 & 0 & 0\\
0 & 0 & 1\\
0 & 1 & 0 \\
\end{pmatrix}.
\end{equation}
The \Z3 PG and SG symmetries acting in \Eqref{eq:PGSG} are the 
subgroups generated by $A^2B^2AB$
and $B$, respectively.
The two-dimensional representations $\rep{2}_1$, $\rep{2}_2$, 
$\rep{2}_3$ and $\rep{2}_4$ are generated by the matrices
\begingroup
\renewcommand*{\arraystretch}{1}
\arraycolsep=3pt
\thinmuskip=1mu
\medmuskip=1mu
\thickmuskip=1mu
\begin{align}\label{eq:DoubletDefs}
\Id_2 = \begin{pmatrix}
1 & 0\\
0 & 1\\
\end{pmatrix},\;\;\;
\Omega_2 = \begin{pmatrix}
\omega^2 & 0 \\
0        & \omega \\
\end{pmatrix}, \;\;\;
S_2 = \begin{pmatrix}
0 & 1\\
1 & 0\\
\end{pmatrix}, \raisetag{20pt}
\end{align}
\endgroup
according to the assignments in \Tabref{tab:DffDoublets}.
\renewcommand{\arraystretch}{1.0}
\begin{table}[t]
\caption{\label{tab:DffDoublets} Explicit matrices for the doublet
representations of $\Delta(54)$, see \eqref{eq:DoubletDefs} for a 
definition of the matrices $\Omega_{\rep{2}}$ and $S_{\rep{2}}$.}
\begin{ruledtabular}
\centerline{\begin{tabular}{ccccc}
   & \rep[_1]{2} & \rep[_2]{2} & \rep[_3]{2} & \rep[_4]{2} \\
   \hline
 $A_{\rep[_i]{2}}$ & $\Id_{\rep{2}}$ & $\Omega_{\rep{2}}$ & $\Omega_{\rep{2}}$ & $\Omega_{\rep{2}}$ \\
 $B_{\rep[_i]{2}}$ & $\Omega_{\rep{2}}$ & $\mathbbm{1}_{\rep{2}}$ & $\Omega_{\rep{2}}$ & $\Omega_{\rep{2}}^*$ \\
 $C_{\rep[_i]{2}}$ & $S_{\rep{2}}$ & $S_{\rep{2}}$ & $S_{\rep{2}}$ & $S_{\rep{2}}$ \\
\end{tabular}}
\end{ruledtabular}
\end{table}\renewcommand{\arraystretch}{1.0}%
In this basis the Clebsch-Gordan coefficients relevant to this work
are given by 
\begingroup
 \arraycolsep=-2pt
 \thinmuskip=0mu
 \medmuskip=0mu
 \thickmuskip=0mu
\newcommand{\hminus}{\hspace{-6pt}}
\begin{subequations}\label{eq:Delta54CGs}
\begin{align}
  \hminus\left(x_{\rep[_i]{2}}\otimes y_{\rep[_i]{2}}\right)_{\rep[_0]{1}} & = \frac{1}{\sqrt{2}}\left(x_1\, y_2 + x_2\, y_1 \right),\\
  \hminus\left(x_{\rep[_i]{3}}\otimes y_{\rep[_i]{\bar{3}}}\right)_{\rep[_0]{1}} & = \frac{1}{\sqrt{3}}\left(x_1\, \bar{y}_1 + x_2\, \bar{y}_2 + x_3\, \bar{y}_3 \right),\\
  \hminus\left(x_{\rep[_i]{3}}\otimes y_{\rep[_i]{\bar{3}}}\right)_{\rep[_1]{2}} & =
  \frac{1}{\sqrt{3}}\left(\begin{array}{c} x_1\, \bar{y}_2 + x_3\, \bar{y}_1 + x_2\, \bar{y}_3 \\ x_2\, \bar{y}_1 + x_1\, \bar{y}_3 + x_3\, \bar{y}_2 \end{array}\right),\\
  \hminus\left(x_{\rep[_i]{3}}\otimes y_{\rep[_i]{\bar{3}}}\right)_{\rep[_2]{2}} & =
  \frac{1}{\sqrt{3}}\left(\begin{array}{c}x_1\, \bar{y}_1 + \omega\, x_2\, \bar{y}_2 + \omega^2\, x_3\, \bar{y}_3 \\ x_1\, \bar{y}_1 + \omega^2\, x_2\, \bar{y}_2 + \omega\, x_3\, \bar{y}_3 \end{array}\right),\\
  \hminus\left(x_{\rep[_i]{3}}\otimes y_{\rep[_i]{\bar{3}}}\right)_{\rep[_3]{2}} & = 
  \frac{1}{\sqrt{3}}\left(\begin{array}{c}x_2\, \bar{y}_3 + \omega\, x_3\, \bar{y}_1 + \omega^2\, x_1\, \bar{y}_2 \\ \omega\, x_2\, \bar{y}_1 + x_3\, \bar{y}_2 + \omega^2\, x_1\, \bar{y}_3 \end{array}\right),\\
  \hminus\left(x_{\rep[_i]{3}}\otimes y_{\rep[_i]{\bar{3}}}\right)_{\rep[_4]{2}} & = 
  \frac{1}{\sqrt{3}}\left(\begin{array}{c}\omega^2\, x_2\, \bar{y}_1 + x_3\, \bar{y}_2 + \omega\, x_1\, \bar{y}_3 \\ x_2\, \bar{y}_3 + \omega^2\, x_3\, \bar{y}_1 + \omega\, x_1\, \bar{y}_2 \end{array}\right).
\end{align}
\end{subequations}
\endgroup

\bibliography{Orbifold}
\bibliographystyle{apsrev}
\end{document}